\let\Xdocument\document
\documentclass{iau}
\let\document\Xdocument

\usepackage{amsmath}
\usepackage{graphicx}
\usepackage{multirow}

\newcommand{\aap}{    {\it A\&A}}

\newcommand{\apj}{    {\it ApJ}}
\newcommand{\apjl}{   {\it ApJL}}
 
\newcommand{\apjs}{    {\it ApJS}}

\newcommand{\solphys}{{\it Solar Phys.}}

\chardef\us=`\_

\begin{document}

\lefttitle{Schmieder et al.}
\righttitle{Rotation of flux ropes in the low corona}

\jnlPage{1}{7}
\jnlDoiYr{2024}
\doival{10.1017/S1743921324001571}

\aopheadtitle{Proceedings IAU Symposium 388}
\editors{N. Gopalswamy,  O. Malandraki, A. Vidotto \&  W. Manchester, eds.}

\title{Rotation of flux ropes in the low corona} 

\author{Brigitte Schmieder $^{1,2}$,  Anwesha Maharana $^1$, Jin Han Guo $^{1,3}$, Luis Linan $^{1}$, Stefaan Poedts $^{1,4}$}



\affiliation{$^1$Centre for mathematical Plasma Astrophysics, Dept. of Mathematics, KU Leuven, 3001 Leuven, Belgium\\
$^2$LIRA, Observatoire de Paris, CNRS, Universit\'e de Paris, 5 place Janssen, 92290 Meudon, France\\
$^3$University  of Nanjing, Nanjing, China\\
$^4$ Institute of Physics, University of Curie-Sk{\l}odowska, Lublin, Poland\\
Email:brigitte.schmieder@obspm.fr}

\begin{abstract}
Eruptions of filaments { are defined by different parameters, specially, sigmoid handedness and  direction of the eruption,  which} are important parameters for forecasting the geoeffectiveness of consequent interplanetary coronal mass ejection (ICME) or magnetic cloud. Solar filaments often exhibit rotation and deflection during eruptions, which would significantly affect the geoeffectiveness of the coronal mass ejections (CMEs). Therefore, understanding the mechanisms that lead to such lateral displacement of filaments is a great concern to space weather forecasting.  Two case studies are discussed.     Firstly, the events of September 8 and September 10 2014, were analyzed from the Sun to the Earth. The numerical heliospheric simulation EUHFORIA shows that the handedness of the EUV sigmoid deduced from coronagraph observations was different from the tilt of the ICME at 1~au, suggesting a rotation of the CME in the low corona. A potential undetected low coronal rotation led to erroneous space weather prediction. The second event concerns a filament observed on August 20 2021, which underwent a rotation of 73 degrees during its eruption, implying a significant lateral drifting of the filament material. A data-constrained magnetohydrodynamic (MHD) simulation confirms such a rotation. 
These two studies reinforce the idea that CMEs are { subjected to more significant rotation and deflection} in the low corona than during their journey in the heliosphere.

\end{abstract}


\maketitle

\section{Introduction}
\label{intro}
Geoeffective event prediction is now possible using MHD simulations (e.g. EUHFORIA). It relies on the orientation of the $B_z$ component of the ejected magnetic flux rope from the Sun, frequently assimilated to a coronal mass ejection (CME). A forecasting limitation arises from the evolution of CME structures during their propagation and their interaction with the solar wind and other CMEs. The interaction of CMEs may lead to severe geoeffective events, as demonstrated by \citet{Shen2018}, \citet{Scolini2020}, and \citet{Koehn2022}. The lack of multiple in situ crossings through CMEs makes predicting its global behaviour through reconstruction using data from a single point challenging. 

A second limitation, such as  CME rotation, deflection, and deformation between their solar source and their consequent ICME at 1au, is even more intriguing. 
Some previous works have indicated that the rotation of CME flux ropes mainly happens in the low corona due to the complicated magnetic field environment. For example, \citet{Shiota2010} found that the rotation of CME flux ropes due to reconnection happens within 5 solar radius.  \citet{Lynch2009} found that the rotation angle of the CME flux rope reaches 50 degrees when it propagates to a distance of 3.5 solar radii. 
{  A rotation up to 56 degrees of a newly built magnetic flux rope by helicity injection was found in the April 23 2023 geo-effective event \citep{Vemareddy2024}}. 
\citet{Kliem2012} highlighted the possibility of extensive low coronal rotation up to even more than 100 degrees by combining twist and shear-driven rotation, the latter being dominant in the lower corona.  MHD simulations from the solar surface to 1 au revealed that the rotation mainly occurs within 15 solar radii \citep{Regnault2023}.
 The boundary conditions of  the European Heliospheric FORecasting Information Asset \citep[EUHFORIA][]{Pomoell2018}  are given at 0.1au and it does not account 
the  CME evolution in the corona below 0.1~au. 
Although the white-light coronagraph images help reconstruct the CMEs in the middle and upper corona, they are insufficient to derive the magnetic field configuration. This leaves us to rely on the source region proxies for guessing the magnetic field configuration for prediction purposes, being agnostic to the dominant low coronal dynamics.\\
Data-driven MHD  simulation seems to be  a good tool to understand better what is happening to a flux rope during its  eruption 
\citep{GuoJH2023a,GuoJH2023b,GuoJH2024}.
In this paper, we present two case studies -- the first paper concerns a complex set of CMEs, showing that EUHFORIA can model the correct magnetic field profiles resulting from the interaction between two CMEs 
by assuming substantial CME rotation in the low corona, 
to justify the consistency of the CME orientation at 0.1 au 
 and at 1~au \citep{Maharana2023}. The second case study shows that data-driven MHD simulation can explain the difference in orientation between the initial eruptive filament and the consequent CME.  In that case, multiple reconnections of flux ropes have been found explaining the observed orientation discrepancy \citep{GuoJH2023b} (Section 3). 
 
 \section{MHD simulation of the September 2014 events}
 \label{September}
 
\begin{figure*}[h!]
\includegraphics[width=0.7\textwidth,angle=-90]{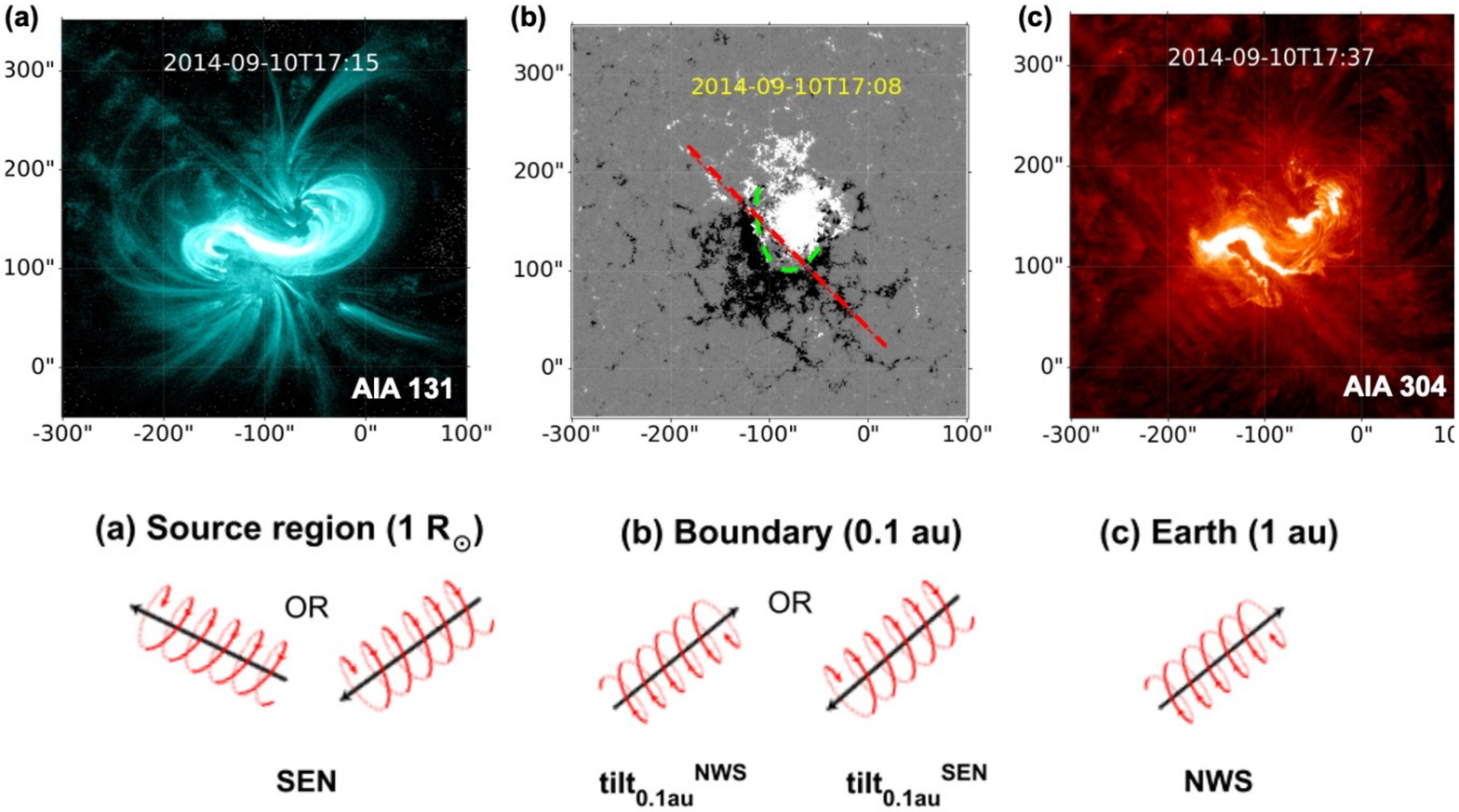}
\caption{Sigmoid observed on September 9 2014, in AR 12158: Top panels: (a)   by AIA in 131 \AA\, (b)  by HMI,  (c) by AIA in 304 \AA.  The red line indicates the tilt of the active region, and the green dashed line is the polarity inversion line. Bottom panels: Schematic representation of the CME2 orientation inferred from different observational proxies at different locations. (a) Close to 1~$R_\odot$, based on the analysis of the source region (sigmoid); (b) close to 0.1~au, based on the 3D reconstruction of the white-light images; and (c) at 1~au, based on the in situ observations. Adapted from \citet{Maharana2023}.}
\label{sigmoid}
\end{figure*}

\begin{figure*}[h!]
\includegraphics[width=1.0\textwidth,trim={1cm 5cm 1cm 5cm},clip=]{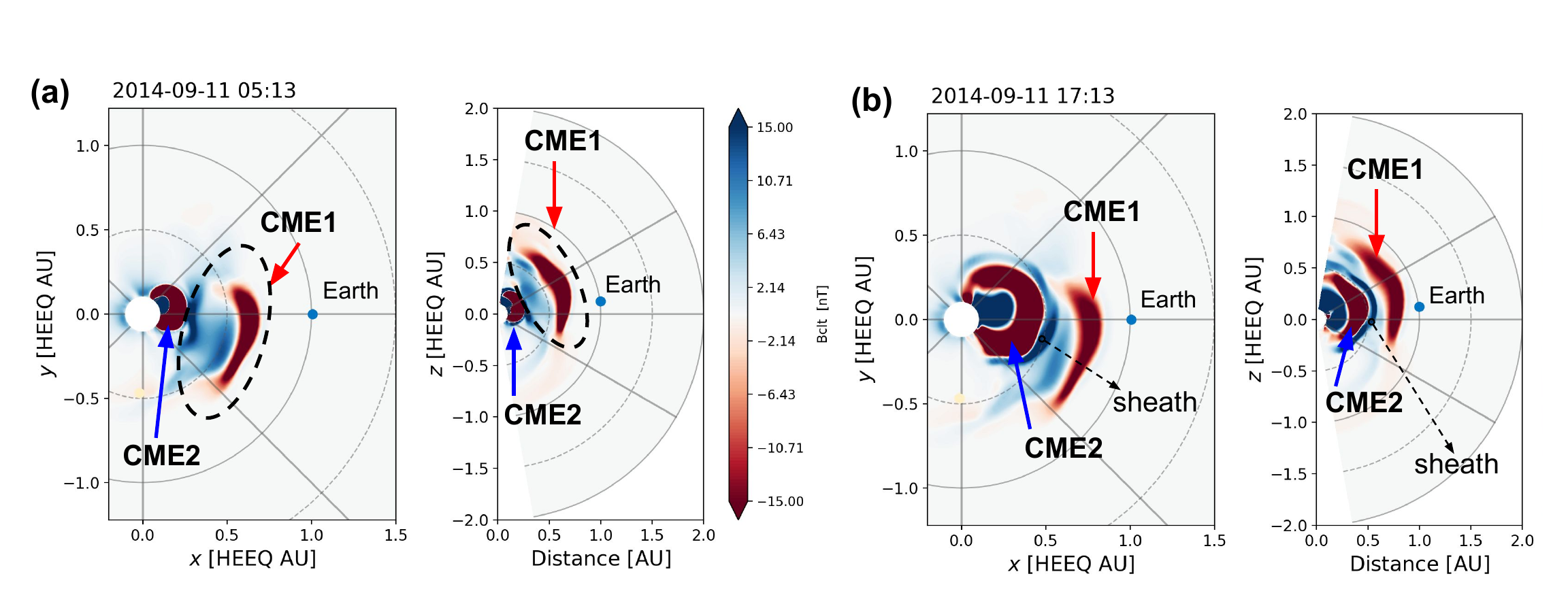}
\caption{Evolution of $B_z$ for the CME-CME interaction event of September 2014 using EUHFORIA. The co-latitudinal component in the spherical coordinate system is $B_{clt}$, equivalent to $-B_z$ on the ecliptic plane. The red and blue spectra of the colour bar correspond to positive and negative $B_z$, respectively. Each sub-figure shows the view of the equatorial (X-Y) and the meridional (X-Z) planes at a particular time mentioned at its top. (a) CME2 is shown evolving behind CME1 in the early stage of interaction; (b) CME2 compresses the trailing negative $B_z$ part of CME1, creating the geo-effective sheath ahead of itself.
Adapted from \citet{Maharana2023}.
}
\label{euhforia}
\end{figure*}

For the first event, we investigate the earthward evolution of two successive CMEs that erupted from the same active region, AR 12158, on September 8, 2014, and September 10, 2014, respectively. This active region was reported to be continuously shearing and deflecting during its evolution \citep{Vemareddy2016}. The first CME (hereafter CME1) was not recorded in any ICME catalogues. However, we noticed a shock signature at Earth and interpreted its atypical magnetic field signatures, which pointed to a potential glancing blow at Earth. 
The second CME (hereafter CME2) was predicted to be a direct hit and cause a major storm due to a strong negative $B_z$ component. This prediction, made by the space weather community, considered only the source region magnetic field signatures of CME2 to interpret the orientation of the incoming flux rope. Surprisingly, a long-lasting positive $B_z$ passed through Earth, with a brief period of negative $B_z$ in the sheath region of the CME that drove the storm. So, instead of a major storm driven by the magnetic cloud, a moderate storm was caused due to the sheath region's magnetic field. Hence, the geoeffectiveness of different CME2 sub-structures was mispredicted. 

The main objective of this study was to understand the change in the CME2 orientation that would have led to the erroneous geoeffectiveness prediction at 1~au. It could be because of the rotation or deformation of CME2 in the lower corona or during its heliospheric propagation. Or it could also be because CME2 impacted Earth with its flank, which gave rise to the in situ magnetic field signatures different from what was predicted, assuming a direct hit based on insights from the early remote-sensing observations. 
We characterised CME2 in three phases, first at the Sun, then in the corona and finally at Earth. We constrained the magnetic axis orientation of the erupting flux rope using the polarity inversion line (PIL) proxy from the HMI magnetogram observations of the active region. The chirality of the flux rope, i.e., the sign of helicity, is determined using the handedness of the EUV sigmoid (Figure~\ref{sigmoid} top panel a). Determining the PIL orientation of this eruption was not straightforward. It was more of a curved topology than a straight line, as shown in Figure~\ref{sigmoid} (top panel b). As our heliospheric simulations require us to adhere to a single orientation, we compromised and chose an average straight-line south-east-north (SEN) configuration as defined by \cite{Bothmer1998}. The schematic representation of the CME front is shown in Figure~\ref{sigmoid} (bottom panel a). We recognized this as one of the potential sources of errors in the prediction. 

Continuing our investigation, we performed a 3D geometric reconstruction of the CMEs in the coronagraph of LASCO and STEREO-B field of view close to 0.1~au. 
The position of CME2 suggested by the 3D reconstruction indicated its deflection in the northwest-ward direction 
making the non-direct encounter at Earth evident. Such reconstructions provide only the geometrical inclination of the flux rope, which opened up two possibilities for the magnetic axis
(Figure \ref{sigmoid} bottom panels). One corresponds to the source region magnetic axis orientation (tilt$^{SEN}$), and the other was rotated more than 200$^{\circ}$ counter-clockwise from the source region orientation (tilt$^{NWS}$, NWS refers to north-west-south).

At 1~au, we performed the numerical fitting of a mathematical flux rope CME model, Flux Rope in 3D \citep[FRi3D,][]{Isavnin2016} to the in situ measurements at Earth using the Flux Rope in 3D \citep[FRi3D][]{Isavnin2016} model. The geometrical orientation of the flux rope obtained from the in situ analysis at 1~au was consistent with tilt$^{NWS}$.

We employed EUHFORIA, a physics-based heliospheric and CME evolution model, to simulate the event. The semi-empirical WSA model obtained the solar wind MHD parameters at the inner boundary. Once a solar wind background was obtained, CMEs were injected as time-dependent boundary conditions. We used the advanced FRi3D model with a global CME geometry to capture the flank encounter of CME2. Finally, the CMEs self-consistently evolved in the solar wind background in the heliospheric domain and forecasting was done. We first performed two simulations just for CME2, initializing it with tilt$^{SEN}$ and tilt$^{NWS}$, respectively. 

With tilt$^{SEN}$, a negative $B_z$ signature was obtained at Earth opposite to what was observed. With tilt$^{NWS}$, prolonged positive $B_z$ was reproduced at Earth. 
Hence, our numerical experiments suggested that low corona rotation was in fact one of the key factors that could not be considered due to the lack of magnetic field observations in the corona. 
After finalizing the CME2 tilt, we perform our final simulation by injecting CME1 before CME2. 
In addition to the long-lasting positive $B_z$ in the CME2 magnetic cloud, we could also model the brief negative $B_z$ in the sheath region formed due to the CME1-CME2 interaction (Figure \ref{September}). The simulation with only CME2 did not result in negative $B_z$ signatures on its front. The sheath region negative $B_z$ was formed due to CME2 compressing and enhancing the negative $B_z$ part of CME1, hence creating the geo-effective sheath. 
This space weather impact of CME-CME interaction would have been difficult to predict just from the CME observations in the corona. Therefore, it is evident that in such cases, we need to combine global modelling of CME propagation in the heliosphere initialized with observations to have an accurate prediction/forecast at 1AU.

In summary, we successfully reproduced the magnetic field components at 1~au for the early September 2014 storm by modelling both CME1 and CME2. We identify the correct orientation of the erupting flux rope at 0.1~au, i.e., the EUHFORIA simulation boundary. It is different from the sigmoid at the source region, and we proposed a possible rotation below 0.1~au, essential for initializing the CME2 in the simulations. We find that the interaction between CME1 and CME2 is responsible for producing the geoeffectiveness in the sheath. Finally, we highlight the importance of 3D MHD modelling of CME evolution in understanding the geoeffectiveness due to the different sub-structures in CME-CME interaction events.

\section{Flux rope rotation in the corona}
\label{Flux_rope}

\begin{figure*}[h!]
\includegraphics[width=16cm]{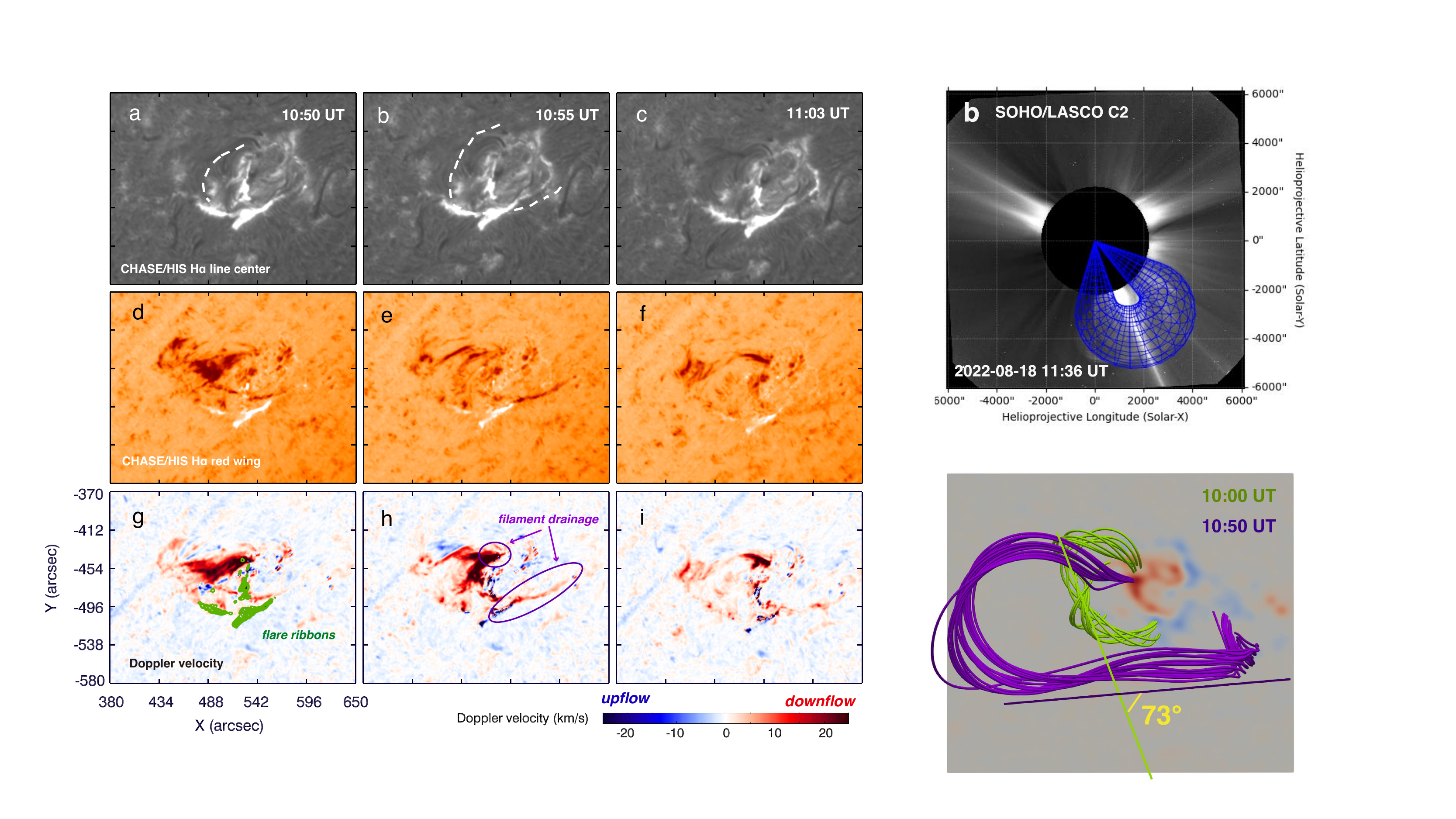}
   \caption{ Observation of the ejection of a filament  and change of orientation by CHASE; top panels: intensity in H$\alpha$,
   middle panels in the H$\alpha$ wing, bottom panels: Doppler shifts. In panels a and b, the spine of the filament is indicated by a dashed line; in panel h, the oval indicates the drainage of the filament towards its drifting footpoint. The CME is on the top right panel, and in the bottom right panel, from the MHD simulation, the initial flux rope is green, and the final flux rope is violet. 73 degrees is the estimated rotation angle of the filament. Adapted from \citet{GuoJH2023b}.}
\label{FR}
\end{figure*}

In the previous section, we show that the orientation of the main flux rope at 1~au 
does not correspond to the source region tilt.  
A rotation of the FR has to be considered in the low corona.
We were tempted to use our data-driven method to see how the sigmoid changed orientation in the low corona. However, this case is very complex and difficult to model, with the interaction of two CMEs and a background solar wind approximated with WSA. We chose a more simple case, the second case.

Solar filaments often exhibit rotation and deflection during eruptions, which would significantly affect the geoeffectiveness of CMEs.  Therefore, understanding the mechanisms that lead to such a rotation and lateral displacement of filaments is a great concern to space weather forecasting.  In fact there are three candidates responsible for the rotation of a CME flux rope in the low corona: (1) kink instability \citep{Rust1996}; (2) lateral Lorentz force induced by sheared/toroidal magnetic fields \citep{Isenberg2007,Kliem2012}; and (3) magnetic reconnection  \citep{Shiota2010,GuoJH2023b}. The rotation due to kink instability mainly occurs in twisted flux ropes. In this scenario, the twist of the eruptive flux rope will release to the right due to helicity conservation. Hence, the rotation direction is consistent with the sign of helicity (positive helicity corresponds to clockwise rotation). As for the effects of sheared/toroidal magnetic fields, they can produce a lateral Lorentz force acting upon the two legs of the flux rope, which can easily lead to large-angle rotation, as validated in Kliem et al. (2012). Unlike the effects of kink instability, the sheared magnetic fields can also result in a large-angle rotation of a sheared arcade, even if the twist number is lower than 1. Regarding the last candidate, the magnetic reconnection between the eruptive flux rope and the ambient sheared arcades can directly change the connectivity of their magnetic field lines, potentially leading to rotation. 
The second event showing a strong rotation of flux rope in the low corona 
belongs to this category.

 We examine an intriguing filament eruption event observed by the Chinese H$\alpha$ Solar Explorer (CHASE) and the Solar Dynamics Observatory (SDO)
on August 18 2022 \citep{GuoJH2023b}. The filament, which evolves into a CME, exhibits significant lateral drifting during its rising. Moreover, the orientation of the CME flux rope axis deviates from that of the pre-eruptive filament observed in the source region. To investigate the physical processes behind these observations, we perform a data-constrained magnetohydrodynamic (MHD) simulation based on MPI-AMRVAC code after inserting a Regularized Biot–Savart laws  (RBSLs) flux rope \citep{Titov2018}. The numerical model reproduces many prominent observational features in the eruption, including the morphology of the eruptive filament, eruption path, and flare ribbons. The simulation results reveal that the magnetic reconnection between the flux rope leg and neighbouring low-lying sheared arcades may be the primary mechanism responsible for the lateral drifting of the filament material (reconnection aa-rf ). Such a reconnection geometry leads to flux rope footpoint migration and a reconfiguration of its morphology. Consequently, the filament material hosted in the flux rope drifts laterally, and the CME flux rope deviates from the pre-eruptive filament (Figure  \ref{FR}). This finding underscores the importance of external magnetic reconnection in influencing the orientation of a flux rope axis during eruption.  The direction of the CME follows the new direction of the FR.  It is interesting to relate the MHD simulation results to the study's conclusion of \cite{Chikunova2023}.
They conclude that the direction of the eruptive filament did not coincide with the main direction of the dimming expansion direction, which reflects the direction of the CME. The simulation and CHASE observations show that the filament had several reconnections before the CME escaped.

\section{Conclusion}
During filament/flux rope eruption, many reconnections with surrounding magnetic fields (arcade top or flux rope leg) may occur as suggested by the 3D standard model \citep{Aulanier2010}. It was demonstrated that in such a model, three types of reconnection with the arcades, with the preexisting flux and in the legs of the flux rope  (aa-fr, ar-fr, fr-fr)  exist \citep{Aulanier2019}.  Such reconnection implies the rotation of the flux rope during its ejection in the low corona. We show two cases: one case where we suspect that there is a rotation of the FR before its journey in the heliosphere using EUHFORIA simulation, and the second case, the effective rotation in the low corona suspected in the CHASE observations and confirmed by an MHD simulation (MPI-AMRVAC). 
{  Rotation of erupting features in the low corona is important for space weather prediction 
\citep{Vemareddy2024}}. 
Therefore, knowledge of the CME magnetic field is crucial for deriving the correct orientation of the {   erupting} flux rope in the low corona and, therefore, for propagating it further in the heliospheric models for space weather forecasting purposes.


\end{document}